\documentclass[prl,superscriptaddress,onecolumnpage,notitlepage]{revtex4-1}

\usepackage{graphicx} 
\usepackage[usenames,dvipsnames]{xcolor}

\usepackage{amsmath}
\usepackage{amssymb}
\usepackage{color}
\usepackage{float}
\usepackage{epstopdf}
\usepackage[normalem]{ulem}
\usepackage{hyperref}
\usepackage{soul}
\usepackage{accents}
\usepackage{pgfplots,pgfplotstable}
\usepackage{soul}

\usepackage{natbib}

\usetikzlibrary[shadings]
\usetikzlibrary{shapes}
\usetikzlibrary{snakes}

\sethlcolor{white}

\newcommand{\vett}[1]{\mathbf{#1}}

\newcommand {\tr} {\mbox{\rm tr\,}}

{\left\lbrace\begin{array}{@{}l@{}}}%
{\end{array}\right.}

\begin{document}
\title{Curvature-Driven Morphing of Non-Euclidean Shells}
\author{Matteo Pezzulla}
\affiliation{
Department of Mechanical Engineering, Boston University, Boston, MA, 02215.
}%

\author{Norbert Stoop}
\affiliation{
Department of Mathematics, Massachusetts Institute of Technology - Cambridge, MA, 02139.
}%

\author{Xin Jiang}
\affiliation{
Department of Mechanical Engineering, Boston University, Boston, MA, 02215.
}%

\author{Douglas P. Holmes}
\email{dpholmes@bu.edu}
\affiliation{
Department of Mechanical Engineering, Boston University, Boston, MA, 02215.
}%

\date{\today}

\begin{abstract}
We investigate how thin structures change their shape in response to non--mechanical stimuli that can be interpreted as variations in the structure's natural curvature. Starting from the theory of non-Euclidean plates and shells, we derive an effective model that reduces a three-dimensional stimulus to the natural fundamental forms of the mid-surface of the structure, incorporating expansion, or growth, in the thickness. Then, we apply the model to a variety of thin bodies, from flat plates to spherical shells, obtaining excellent agreement between theory and numerics. We show how cylinders and cones can either bend more or unroll, and eventually snap and rotate. We also study the nearly-isometric deformations of a spherical shell and describe how this shape change is ruled by the geometry of a spindle. As the derived results stem from a purely geometrical model, they are general and scalable.
\end{abstract}

\pgfplotsset{compat=newest}

\maketitle

\section{Introduction}

Non--mechanical stimuli, such as temperature, pH, and swelling have significant influences on the shape of a structure. The bending of bimetal thermostats is perhaps the first rationalized example of how temperature can induce shape changes~\cite{Timoshenko1925}. Under a homogeneous temperature change, the two metal layers in the bimetal expand by different amounts, resulting in a uniform bending deformation with residual thermal stresses~\cite{Ozakin2010}. The swelling of gels, elastomers, and foams can cause a significant increase in their volume~\cite{Hong2008,Lucantonio2013}, and if swelling causes different parts of a material to expand more in response to the same amount of solvent, nontrivial shape changes can occur~\cite{Klein2007,Holmes2011,Kim2012a}. The physics behind thermal expansion and swelling are rather different from each other, yet the resulting shape changes can sometimes be qualitatively and quantitatively similar, highlighting the underlying geometric connection of these non--mechanical stimuli to local volume changes in the material. If the structure is thin, even small volume changes can have dramatic consequences on an object's shape. These large, geometrically nonlinear deformations are ever present in the growth and reconfiguration of thin biological structures. For instance, the contraction of the spherical, fluid--filled shell of a {\em Volvox} embryo turns itself inside out to enable motility~\cite{Hohn2015,Haas2015}, while the differential drying of a pollen grain causes it to morph into a spindle to slow dessication~\cite{Katifori2010}. Differential growth rates in the plane of a growing leaf will cause it to develop wrinkles along its edge~\cite{Boudaoud2010}. These morphogenetic shapes have analogues with completely mechanical systems as well, where pre-stretch can play the role of growth. Although swelling may develop in a variety of ways, it is usually convenient to identify two main categories, which are in-plane or through-the-thickness differential swelling. When swelling develops within the mid-surface of the body, flat plates can buckle into hyperbolic surfaces~\cite{Klein2011,Pezzulla2015} or bend into domes~\cite{Kim2012a}, and when swelling develops across the thickness, plates can deform into cylinders~\cite{Pezzulla2016}.  

The modeling of growth in three-dimensional elasticity has led to the so-called \emph{incompatible elasticity}, which began with the introduction of the multiplicative decomposition of the deformation gradient in~\cite{Kondaurov1987,Takamizawa1987}, borrowed from models in plasticity. Since thin structures can undergo small stretching strains as a consequence of their slenderness~\cite{Rayleigh1894}, scientists developed reduced ordered models in complete analogy to what has been done in standard mechanics for the theory of plates and shells~\cite{Love1888,VonKarman1910,Koiter1973}. In particular, a theory for the growth of plates and shells in the context of incompatible elasticity, named theory of non-Euclidean plates and shells, was presented in~\cite{Efrati2009} and successfully applied in~\cite{Armon2011,Pezzulla2015,Pezzulla2016}, just to name a few. A refined geometric theory of morphoelastic shells was also recently derived in~\cite{Sadik2016}. 
The large deformability of thin structures offers a pathway towards the design of smart systems that can completely change their shape~\cite{Py2007}, while at times taking advantage of elastic instabilities~\cite{AmarGoriely2005,Goriely2005}. This connection between differential expansion, or growth, to thin, naturally curved structures motivates the study of how shells deform in response to variations in their natural curvature.

In this paper, we investigate the shape changes that can be induced on plates and shells when a non--mechanical stimulus acts through-the-thickness. Starting from the theory of non-Euclidean plates and shells, we derive an expression for the natural forms of the shell, which also takes into account the possible conformal growth of the mid-surface. Although we apply the model to study bilayer shells, the presented theory effectively does not require a physical bilayer material, as long as the stimulus has a gradient along the thickness. As the non--mechanical stimulus is enclosed in the concept of natural curvature~\cite{Audoly2010}, the derived models are general, purely geometrical and, as such, scalable. The paper is organized as follows. We initially set the notation and provide a brief summary of the theory of non-Euclidean plates and shells, and then present a straightforward procedure to derive the expressions of the natural forms when a general plate or a shell is subjected to a stimulus that can be represented by a variation in its natural curvatures. We show how the resulting natural curvature can be seen as the sum of a homothety of the initial shape plus a spherical shape--changing natural curvature. We apply this general method to intrinsically flat shells such as plates, cylinders, and cones. When the shape is extrinsically curved, we also show how growth can trigger snap-through instabilities, and provide a simple formula for the stability threshold in terms of the natural curvature. Finally, we study the effects of variations in natural curvature of spherical shells, validating the presented model through a comparison of 3D and 2D numerical results. Moreover, we show how this shape change is similar to the one developed by pollen grains during harmomegathy~\cite{Katifori2010}, and show how these shapes are similar to spindles.

\section{General considerations}
\label{gencons}

\begin{figure}[t]
\centering
\includegraphics[scale=1]{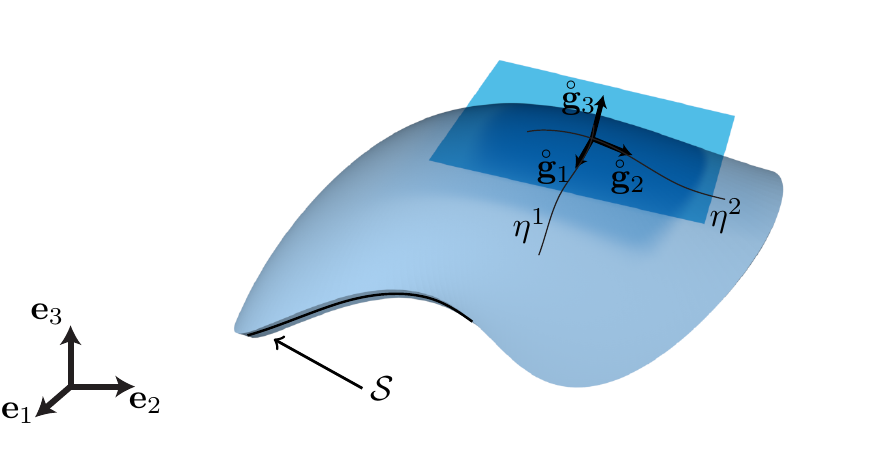}

\caption{A shell in its reference state embedded in the Euclidean space. The tangent coordinates~$(\eta^1,\eta^2)$ are shown as well as the covariant base vectors.\label{scheme}}

\end{figure}

In this section, we recall some standard concepts in shell theory, set the notation that will be used throughout the paper, and summarize the theory of non-Euclidean shells when the growth along the thickness is taken into account.
We identify a shell with its mid-surface~$\mathcal{S}\subset\mathcal{E}$, where~$\mathcal{E}$ is the three-dimensional Euclidean space provided with a cartesian basis~$(\vett{e}_1,\vett{e}_2,\vett{e}_3)$ as shown in Figure~\ref{scheme}. The shell, viewed as a three-dimensional body, is a bounded domain~$\mathcal{B}\subset\mathcal{E}$, such that~$\mathcal{B}=\mathcal{S}\times[-h/2,h/2]$, where~$h$ is its thickness. The body can be naturally endowed with a system of curvilinear coordinates chosen in reference to the embedding of the initial configuration~$\accentset{\circ}{\vett{r}}(\eta^1,\eta^2,\eta^3)\colon\mathcal{D}\rightarrow\mathcal{E}$, where we choose~$(\eta^1,\eta^2)$ so that they span the mid-surface of the shell while~$\eta^3$ runs along the normal~$\accentset{\circ}{\vett{n}}$ to the mid-surface, and~$\mathcal{D}$ represents the domain of parametrization of the reference configuration. In the following, we will denote with~$\vett{r}$ the embedding of the deformed configuration, and with~$\vett{n}$ the normal to the deformed mid-surface. Not only does the parametrization of the initial configuration endow the body with a system of coordinates, but it also provides it with a three-dimensional metric~$\accentset{\circ}{\vett{g}}$ and a covariant basis~$(\accentset{\circ}{\vett{g}}_1,\accentset{\circ}{\vett{g}}_2,\accentset{\circ}{\vett{g}}_3)$. The covariant metric coefficients of the reference configuration are defined as~$\accentset{\circ}{g}_{ij}=\accentset{\circ}{\vett{r}},_i\cdot\accentset{\circ}{\vett{r}},_j$, where commas denote partial derivatives, Latin indices run from~$1$ to~$3$, and the symbol~$\cdot$ denotes the standard inner product in the Euclidean space. When one deals with standard compatible elasticity, meaning that no inelastic stimuli act on the body, the metric~$\accentset{\circ}{\vett{g}}$ is the base state for the measurement of strains. However, when the body is subjected to some inelastic stimuli such as growth, the strain is measured with respect to a relaxed, natural metric~$\vett{\bar{g}}$, modeling the rest lengths induced by the stimulus~\cite{AmarGoriely2005,Nardinocchi2007,Efrati2009,Yavari2010}. While the metric~$\accentset{\circ}{\vett{g}}$ is derived from the embedding~$\accentset{\circ}{\vett{r}}$, the natural metric~$\bar{\vett{g}}$ is formulated {\it ad hoc} for the specific stimulus at hand. As a result, the metric~$\accentset{\circ}{\vett{g}}$ trivially corresponds to a null Riemann curvature tensor while the natural metric~$\bar{\vett{g}}$ may not, in general, and cannot be embedded in the Euclidean space~\cite{Oneill1997}. Moreover, we denote with~$g_{ij}=\vett{r},_i\cdot\vett{r},_j$ the covariant metric coefficients of the deformed configuration. Finally, as the time scales associated with growth or swelling are much larger than those associated with inertia, we neglect inertia loads and, in general, the following analyses hold for quasi-static stimuli.

Since reduced order models are based on the geometry of the mid-surface, we briefly recall some concepts from differential geometry following~\cite{Oneill1997}. The first fundamental form~$\accentset{\circ}{\vett{a}}$ of a surface parametrized by~$\accentset{\circ}{\vett{R}}=\accentset{\circ}{\vett{r}}\lvert_{\eta^3=0}$ has covariant components~$\accentset{\circ}{a}_{\alpha\beta}=\accentset{\circ}{\vett{R}},_\alpha\cdot\accentset{\circ}{\vett{R}},_\beta$ while the second fundamental form~$\accentset{\circ}{\vett{b}}$ has covariant components~$\accentset{\circ}{b}_{\alpha\beta}=\accentset{\circ}{\vett{n}}\cdot\accentset{\circ}{\vett{R}},_{\alpha\beta}$ (Greek indices run from~$1$ to~$2$). The eigenvalues of the second fundamental form are the principal curvatures of the surface: the average of the eigenvalues is called mean curvature~$H$, and their product is the Gaussian curvature~$K$. Probably the most important theorem in differential geometry, Gauss's Theorema Egregium, states that the Gaussian curvature is an isometric invariant. Finally, we denote with~$a_{\alpha\beta}=\vett{R},_\alpha\cdot\vett{R},_\beta$ and~$b_{\alpha\beta}=\vett{n}\cdot\vett{R},_{\alpha\beta}$, the first and second fundamental forms of the deformed mid-surface, respectively.
The theory is derived by assuming a plane state of stress first, followed by considering Kirchhoff-Love kinematics~\cite{Efrati2009}. The Kirchhoff-Love assumption can be stated as~$\varepsilon_{\alpha3}=0$, where~$2\varepsilon_{ij}=g_{ij}-\bar{g}_{ij}$ is the strain defined as the difference between the visible and the natural metric. On the other hand, if~$\vett{S}$ and~$\vett{T}$ denote the first Piola-Kirchhoff and Cauchy stress tensors, respectively, the plane stress condition may be stated equivalently either as~$\vett{S}\accentset{\circ}{\vett{n}}=\vett{0}$ or~$\vett{T}\vett{n}=\vett{0}$.
In~\cite{Efrati2009}, the stimulus is assumed to be planar, \textit{i.e.} tangent to the mid-surface at each point, thus not inducing a change of the rest length along the thickness. This is usually stated as~$\bar{g}_{i3}=\delta_{i3}$ that implies, through the plane stress assumption, $g_{i3}=(1-\nu/(1-\nu)\varepsilon_{\alpha\alpha})\delta_{i3}$.
The dimensionless elastic energy is usually written as~\cite{Efrati2009}
 
\begin{equation}\label{eq:energy}
\overline{\mathcal{U}}=\int_{\omega} [(1-\nu)\tr(\vett{a}-\vett{\bar{a}})^2+\nu\tr^2(\vett{a}-\vett{\bar{a}})]d\omega+\frac{h^2}{3}\int_{\omega} [(1-\nu)\tr(\vett{b}-\vett{\bar{b}})^2+\nu\tr^2(\vett{b}-\vett{\bar{b}})]d\omega\,,
\end{equation}
where~$\nu$ is the Poisson ratio, $\tr$ denotes the trace operator in the surface metric defined by~$\bar{\vett{a}}$, $d\omega$ is the relaxed area element~$d\omega=\sqrt{\det\bar{\vett{a}}}\ d\eta^1d\eta^2$, $\vett{a}$ is the first fundamental form of the mid-surface, containing all information about lateral distances between points, and~$\vett{b}$ is the second fundamental form of the mid-surface, containing all information about the local curvature. The natural first and second fundamental forms~$\vett{\bar{a}}$ and~$\vett{\bar{b}}$ represent the lateral distances and curvatures that would make the sheet locally stress-free, and they are determined by the specific stimulus. As forms of a surface embedded in Euclidean space, $\vett{a}$ and~$\vett{b}$ must satisfy the Gauss-Codazzi-Mainardi equations~\cite{Oneill1997}, while the natural forms~$\vett{\bar{a}}$ and~$\vett{\bar{b}}$ do not have the same constraints, and can be incompatible. This is analogous to the lack of a three-dimensional compatibility requirement on~$\bar{\vett{g}}$, which could be not embeddable in the Euclidean space, differently from~$\vett{g}$. It is this incompatibility, in addition to that imposed by geometric confinement when present, which drives the deformation of the shell. 

When the stimulus is spherical, as for swelling, $\bar{g}_{i3}=\Lambda_\textup{o}^2\delta_{i3}$. In this case, the plane stress assumption implies~$g_{i3}=(\Lambda_\textup{o}^2-\nu/(1-\nu)\varepsilon_{\alpha\alpha})\delta_{i3}$~\cite{Efrati2010}~\footnote{We are not aware of a published version of this paper other than an electronic version at the link provided in the bibliography.}. 
This means that the three-dimensional metric can be expanded through the thickness up to the first order as~$\vett{g}=\vett{a}-2\eta^3|\vett{r},_3|\vett{b}$. By recognizing that~$|\vett{r},_3|=\sqrt{g_{33}}=\Lambda_o+O(\varepsilon)$, the expansion reads
\begin{equation}\label{exp}
\vett{g}=\vett{a}-2\eta^3\Lambda_\textup{o}\vett{b}\,,
\end{equation}
where the first and second fundamental forms can be computed as
\begin{equation}\label{ab}
\vett{a}=\vett{g}|_{\eta^3=0}\,,\quad\vett{b}=-\frac{1}{2}g^{33}\frac{\partial\vett{g}}{\partial\eta^3}\Bigr|_{\eta^3=0}=-\frac{1}{2\Lambda_o}\frac{\partial\vett{g}}{\partial\eta^3}\Bigr|_{\eta^3=0}\,.
\end{equation}
The expansion resulting in equation~\eqref{exp} differs from the one presented in~\cite{Efrati2009} by the factor~$\sqrt{g_{33}}\simeq\Lambda_\textup{o}$ that multiplies the thickness coordinate~$\eta^3$~\cite{Efrati2010}.  This results from the variation in length along the thickness, which is taken into account.
Let us also note that the expansion~\eqref{exp} leads to a slightly different expression of the energy of the non-Euclidean shell:
\begin{equation}\label{newenergy}
\overline{\mathcal{U}}=\int_\omega [(1-\nu)\tr(\vett{a}-\vett{\bar{a}})^2+\nu\tr^2(\vett{a}-\vett{\bar{a}})]d\omega+\Lambda_\textup{o}^2\frac{h^2}{3}\int_\omega [(1-\nu)\tr(\vett{b}-\vett{\bar{b}})^2+\nu\tr^2(\vett{b}-\vett{\bar{b}})]d\omega\,,
\end{equation}
which takes into account the growth of the thickness; note the emergence of the $\Lambda_\textup{o}^2$ term that multiplies the bending energy~\cite{Efrati2010}. We recall that this model is suitable for the description of the large displacement but small strain deformations of a shell. When large strains are involved, dimensional reduction should be carried out by starting from other hyperelastic models, such as Mooney-Rivlin or neo-Hooke~\cite{Lucantonio2016a}.

\section{Natural forms of shells} 
\label{NatForms}

As a stimulus is intrinsically three-dimensional, the natural metric~$\bar{\vett{g}}$ that describes it has to be consistently reduced to the first and second natural fundamental forms~$\bar{\vett{a}}$ and~$\bar{\vett{b}}$ of the mid-surface. One example of the dimensional reduction of a three-dimensional natural metric into first and second natural fundamental forms was nicely presented in~\cite{Armon2011}. In that case, two flat sheets were pre-stretched by different amounts in different directions and then glued on top of each other. The pre-stretch of the sheet is viewed as an in-plane conformal stretch of the three-dimensional reference metric of each sheet, neglecting the Poisson effect along the thickness. If the two sheets are pre-stretched differently (in magnitude and/or direction), the natural metric of the bilayer sheet is discontinuous. When a metric discontinuity is encountered along the mid-surface of the sheet, it cannot be approximated as continuous, and bending energy will smooth out the discontinuity in the realized, visible metric as well documented in~\cite{Moshe2013}, and shown in~\cite{Pezzulla2015}. However, when the discontinuity develops across the thickness of the sheet, \textit{i.e.} the smallest length of the body, the natural metric may be approximated as a continuous linear function of~$\eta^3$~\cite{Armon2011}. This approximation is consistent with the theory of plates and shells, as the three-dimensional visible metric is expanded linearly along the thickness~\cite{Koiter1973,Ciarlet2000,Efrati2009}. 

In~\cite{Armon2011}, only initially flat sheets were considered, neglecting the influence on initial curvature and the conformal stretching of the mid-surface. Moreover, the approach is based on a linear approximation of the target metric without a geometrical criterion, and therefore provides only a qualitative result. In this Section, we present a quantitative, formal derivation of the first and second natural fundamental forms in the general case of natural stretch and curvature of an initially curved sheet, where the reference three-dimensional metric is a general function of~$\eta^3$. Then, we specialize this result to the case of bilayer shells.

Let us consider a shell with initial, first and second fundamental forms~$\accentset{\circ}{\vett{a}}$ and~$\accentset{\circ}{\vett{b}}$ (figure~\ref{scheme}). We assume that the reference three-dimensional metric is a general function of~$\eta^3$ and pose the equivalence problem of finding the corresponding natural first and second natural fundamental forms. We recall that the reference metric~$\bar{\vett{g}}(\eta^\alpha,\eta^3)$ is spherical and can be written as~$\bar{\vett{g}}(\eta^\alpha,\eta^3)=f(\eta^3)\bar{\vett{g}}(\eta^\alpha)$, that is each component of the tensor depends on~$\eta^3$ via the function~$f$. Within a first-order theory of shells, it is consistent to approximate any function of the thickness coordinate as a linear function, and therefore we proceed to determine the projection of~$f(\eta^3)$ into the subspace of all affine functions in~$[-h/2,h/2]$, of which ~$\{\sqrt{1/h},\sqrt{12/h^3}\eta^3\}$ represents an orthonormal basis. The linear projection of~$\bar{\vett{g}}(\eta^\alpha,\eta^3)$ can be then written as
\begin{equation}
\bar{\vett{g}}^\textup{p}(\eta^\alpha,\eta^3)=\frac{1}{h}\int\bar{\vett{g}}(\eta^\alpha,\eta^3)1d\eta^3+\frac{12}{h^3}\eta^3\int\bar{\vett{g}}(\eta^\alpha,\eta^3)\eta^3d\eta^3\,.
\end{equation}
To keep the notation simple, we will disregard the suffix~\emph{p} hereafter. To determine the natural first and second fundamental forms induced on the mid-surface by the three-dimensional reference metric, we use~\eqref{ab} for the natural metric and find:
\begin{equation}\label{abar}
\bar{\vett{a}}=\vett{\bar{g}}|_{\eta^3=0}=\frac{1}{h}\int\bar{\vett{g}}(\eta^\alpha,\eta^3)1d\eta^3\,,
\end{equation}     
\begin{equation}\label{bbar}
\bar{\vett{b}}=-\frac{1}{2}\bar{g}^{33}\frac{\partial\vett{\bar{g}}}{\partial\eta^3}\Bigr|_{\eta^3=0}=-\frac{6}{h^3}\bar{g}^{33}\int\bar{\vett{g}}(\eta^\alpha,\eta^3)\eta^3d\eta^3\,.
\end{equation}

When the shell is made of two layers of thicknesses~$h_\textup{outer}$ and~$h_\textup{inner}$ respectively, we can assume that the outer layer is locally swelling by a factor~$\alpha^2$ whereas the inner layer is locally swelling by a factor~$\beta^2$ (the function~$f(\eta^3)$ is discontinuous). Rest lengths and volumes can be identified by the three-dimensional natural metric within each layer as~$\overline{\vett{g}}=\alpha^2\accentset{\circ}{\vett{g}}$ and~$\overline{\vett{g}}=\beta^2\accentset{\circ}{\vett{g}}$ in the outer and inner layer, respectively. Equations~\eqref{abar} and~\eqref{bbar} become
\begin{equation}
\bar{\vett{a}}=\Bigl(\frac{m}{1+m}\alpha^2+\frac{1}{1+m}\beta^2\Bigr)\accentset{\circ}{\vett{g}}(\eta^1,\eta^2,0)=:\Lambda_\textup{o}^2\accentset{\circ}{\vett{a}}(\eta^\alpha)\,,
\end{equation}     
\begin{equation}\label{bbarnew}
\bar{\vett{b}}=\Lambda_\textup{o}\accentset{\circ}{\vett{b}}(\eta^\alpha)-\frac{1}{\Lambda_\textup{o}}\frac{3}{h}\frac{m}{(1+m)^2}(\alpha^2-\beta^2)\accentset{\circ}{\vett{a}}(\eta^\alpha)\,,
\end{equation}
where~$m=h_\textup{outer}/h_\textup{inner}$. Before proceeding further, notice that in the case of homogenous stimulus (heating, swelling, etc), $\alpha=\beta$ and the second addend in the expression of the natural second fundamental form is equal to zero, meaning that there is no differential swelling and the shell will just increase its dimensions homothetically as we will discuss later in more detail. 

An important concept related to the natural forms is that of natural curvature~\cite{Audoly2010}, since it has a direct and visible meaning: it is the curvature of a beam cut out of the shell. Being a one-dimensional object, the beam is able to adopt the natural curvature that alludes the shell due to constraints placed on the geometry of surfaces. In general, the natural curvature varies with the direction of the cut and the point in which it is evaluated. Let us recall that the principal curvatures~$\kappa_\delta$ of a surface with first fundamental form~$\vett{a}$ and second fundamental form~$\vett{b}$ can be computed by solving the eigenvalue problem~$\det(\vett{b}-\kappa_\delta\vett{a})=0$. Although~$\bar{\vett{a}}$ and~$\bar{\vett{b}}$ may be incompatible, \textit{i.e.} they do not satisfy the Gauss-Codazzi-Mainardi equations, and therefore they are not the fundamental forms of a surface in the physical Euclidean space, they correspond to a natural surface embedded in a curved space. Therefore, the principal natural curvatures~$\bar{\kappa}_\delta$ may be computed by solving the eigenvalue problem

\begin{equation}\label{detbar}
\det(\bar{\vett{b}}-\bar{\kappa}_\delta\bar{\vett{a}})=0\,.
\end{equation}
Since we can write equation~\eqref{bbar} in mixed coordinates as

\begin{equation}
\bar{b}_\alpha^{\gamma}=\frac{1}{\Lambda_\textup{o}}\accentset{\circ}{b}_\alpha^{\gamma}-\frac{1}{\Lambda_\textup{o}^3}\frac{3}{h}\frac{m}{(1+m)^2}(\alpha^2-\beta^2)\delta_{\alpha}^{\gamma}\,,
\end{equation}
equation~\eqref{detbar} implies
\begin{equation}\label{kalpha}
\bar{\kappa}_\delta=\frac{\accentset{\circ}{\kappa}_\delta}{\Lambda_\textup{o}}-\frac{1}{\Lambda_\textup{o}^3}\frac{3}{h}\frac{m}{(1+m)^2}(\alpha^2-\beta^2)=:\frac{\accentset{\circ}{\kappa}_\delta}{\Lambda_\textup{o}}+\bar{\kappa}\,,
\end{equation}
where~$\accentset{\circ}{\kappa}_\delta$ are the principal curvatures of the mid-surface in the reference configuration. As~$\bar{b}_{\alpha\beta}=\bar{b}_{\alpha}^{\gamma}\bar{a}_{\gamma\beta}$, the second natural fundamental form can be written in terms of the first natural fundamental form and natural curvatures. Notice that differential swelling (or residual swelling, heating, etc) is additive to the initial geometric curvature except for the prefactor~$1/\Lambda_\textup{o}$, which is fundamental for the description of homotheties. This also means that if a straight bilayer beam bends and achieves a curvature~$\bar{\kappa}$, a beam that has an initial curvature~$1/R$ and the same thickness structure will simply achieve a final curvature~$1/(\Lambda_\textup{o}R)+\bar{\kappa}$. As regards residual swelling~\cite{Pezzulla2015}, $\Lambda_\textup{o}\simeq1$ and so the additive decomposition of curvatures becomes even more evident.

The main result of this section is that the stimulus affects the natural forms in two ways: a conformal stretch of the mid-surface and a variation of the curvature tensor. Equation~\eqref{kalpha} provides a simple quantitative prediction of the natural curvature induced by a three-dimensional stimulus, based on the linear projection of the reference three-dimensional metric. The prediction for~$\bar{\kappa}$, namely
\begin{equation}\label{kappa}
\bar{\kappa}=-\frac{1}{\Lambda_\textup{o}^3}\frac{3}{h}\frac{m}{(1+m)^2}(\alpha^2-\beta^2)\,,
\end{equation}
is exactly equal to the curvature of bilayered beams presented in~\cite{Lucantonio2014a}, for small strains and homogenous material. This is remarkable since equation~\eqref{kappa} derives from purely geometrical principles and the projection of the three-dimensional reference metric into the subspace of linear functions of the thickness coordinate. In the remaining part of the paper, the conformal stretching factor~$\Lambda_\textup{o}$ and the natural curvature~$\bar{\kappa}$ will be used as input parameters of the models, as they are directly experimentally measurable in contrast to~$\alpha,\ \beta$ and~$m$. Equation~\eqref{kappa}, or the more general one deriving from equations~\eqref{abar} and~\eqref{bbar}, can be used to obtain a quantitative simple relation between these input parameters and the three-dimensional stimulus.

\subsection{Homogenous heating or swelling of a spherical shell}

The problem of the homogeneous heating or swelling of a spherical shell is a clear example in which one has to take into account the change of the rest length along the thickness of the shell, which would otherwise be constrained to grow only along the mid-surface keeping its radius constant. The homogenous growth takes place at zero elastic cost, meaning that both stretching and bending energies are zero. As this requires the natural forms to be compatible, it is interesting to investigate under which conditions the natural forms derived in this section can be compatible in the case of a spherical shell. So, let us write the Gauss-Codazzi-Mainardi equations for orthogonal natural forms:

\begin{equation}\label{GCM}
\begin{aligned}
&\bar{b}_{11},_2=\bar{b}_{11}\bar{\Gamma}_{12}^1-\bar{b}_{22}\bar{\Gamma}_{11}^2\,,\\
-&\bar{b}_{22},_1=\bar{b}_{11}\bar{\Gamma}_{22}^1-\bar{b}_{22}\bar{\Gamma}_{12}^2\,,\\
&\bar{K}=-\frac{1}{\sqrt{\bar{a}_{11}\bar{a}_{22}}}\biggl[\biggl(\frac{(\sqrt{\bar{a}_{22}}),_1}{\sqrt{\bar{a}_{11}}}\biggr),_1+\biggl(\frac{(\sqrt{\bar{a}_{11}}),_2}{\sqrt{\bar{a}_{22}}}\biggr),_2\biggr]\,,\\
\end{aligned}
\end{equation}
where~$\bar{\Gamma}_{\alpha\beta}^{\delta}$ denotes the Christoffel symbol of the second kind associated with the metric~$\bar{\vett{a}}$.
The first two equations are the Codazzi-Mainardi equations, which represent a structural condition on the second derivatives of the Gauss map, while the last equation represents Gauss's Theorema Egregium for orthogonal metrics~\cite{Oneill1997}. For the problem at hand, $\bar{\vett{a}}=\Lambda_\textup{o}^2\accentset{\circ}{\vett{a}}$ and~$\bar{\vett{b}}=\Lambda_\textup{o}\accentset{\circ}{\vett{b}}$, where~$\accentset{\circ}{\vett{a}}$ and~$\accentset{\circ}{\vett{b}}$ are the first and second fundamental form of the undeformed mid-surface of the shell. Since~$\accentset{\circ}{\vett{a}}$ and~$\accentset{\circ}{\vett{b}}$ satisfy the Gauss-Codazzi-Mainardi equations by definition, we easily see that also the scalar multiples~$\bar{\vett{a}}$ and~$\bar{\vett{b}}$ do. Therefore, the natural forms are compatible and homogenous swelling or growth can take place at zero elastic cost. As a result, a spherical shell of radius~$R$ will grow to a spherical shell of radius~$\Lambda_\textup{o}R$. If growth along the thickness was not accounted for, the radius would be constrained to stay constant, being the growth limited to the tangent space. 

Let us now consider the more general case of differential swelling of a spherical shell of radius~$R$. In this case, equations~\eqref{abar} and~\eqref{bbar} can be written as~$\bar{\vett{a}}=\Lambda_\textup{o}^2\accentset{\circ}{\vett{a}}$ and~$\bar{\vett{b}}=\kappa_\textup{o}\bar{\vett{a}}$. Substituting these expressions in equations~\eqref{GCM}, we find that the Codazzi-Mainardi equations are still satisfied while the Gauss equation is verified if and only if~$\kappa_\textup{o}=1/(\Lambda_\textup{o}R)$, so only in the homogenous swelling case. When growth is triggered by variations in natural curvature and conformal stretch of the mid-surface, we can then conclude that incompatibility arises only from the Gauss's Theorema Egregium.

\section{Intrinsically flat shells}
\label{IntFlat}

To test this model on differential swelling or growth, we start with the bilayer growth of intrinsically flat shells such as plates, cylinders and cones. Eventually, we will show how plates and cylinders can be recovered as particular cones. 

\subsection{Plates}

The case of plates in which one layer swells relative to the other was largely studied in~\cite{Pezzulla2016}, where the main result was that incompressible plates eventually bend into cylindrical shapes whose mean curvature is three-fourths the natural curvature divided by the square of the conformal stretch factor~$H=(3/4)\kappa_\textup{o}/\Lambda_\textup{o}^2$. This result stemmed from first and second natural fundamental forms written as~$\bar{\vett{a}}=\Lambda_\textup{o}^2\vett{I}$ and~$\bar{\vett{b}}=\kappa_\textup{o}\vett{I}$, while their correct expressions were derived in the previous section as (for the case of plates)~$\bar{\vett{a}}=\Lambda_\textup{o}^2\vett{I}$ and~$\bar{\vett{b}}=\kappa_\textup{o}\Lambda_\textup{o}^2\vett{I}$. This results in a formula for the mean curvature that does not depend on the conformal stretch factor, namely
\begin{equation}\label{34plate}
H=\frac{3}{4}\kappa_\textup{o}\,.
\end{equation} 
While this does not affect the results presented in~\cite{Pezzulla2016} as~$\Lambda_\textup{o}\simeq1$ in that case, it can have large effects for high swelling ratios. We will recover the case of flat plates from that of cylinders presented in the next section.

\subsection{Cylinders} Let us consider an open bilayer cylindrical shell of radius~$R$ in which one layer shrinks and the other layer swells. If the swelling layer is on the outer part of the cylinder, the shell will increase its curvature; on the contrary, if the inner layer swells, the cylinder will unroll and eventually snap-through, and bend along a direction orthogonal to the initial one, similar to snap bracelets~\cite{Kebadze2004}. As the shell is thin, it will try to minimize its stretching energy as much as possible so that in the isometric limit it will be exactly equal to zero. For finite thicknesses, the deformation will slightly deviate from the isometric limit due to the presence of boundary layers of width~$\sqrt{h/\bar{\kappa}}$. For a cylinder with cylindrical coordinates~$(\eta^1,\eta^2)=(\theta,z)$ we have from equation~\eqref{bbarnew}:
\begin{equation}
\bar{a}_{\alpha\beta}=\Lambda_\textup{o}^2\accentset{\circ}{a}_{\alpha\beta}=
\Lambda_\textup{o}^2
\begin{pmatrix}
R^2&0\\
0&1\\
\end{pmatrix}\,,\quad
\bar{b}_\alpha^{\eta}=
\begin{pmatrix}
\frac{1}{\Lambda_\textup{o}R}+\bar{\kappa}&0\\
0&\bar{\kappa}\\
\end{pmatrix}\,.
\end{equation}

To determine the shape of the cylinder that corresponds to a particular value of the natural curvature, we minimize the bending energy given by the second addend in equation~\eqref{newenergy} under the isometric constraint~\cite{Pezzulla2016}. We emphasize that for a flat homogeneous metric, as in the case of a cylinder, the Gauss-Codazzi-Mainardi equations admit homogenous second fundamental forms as solutions. Consequently, when minimizing the bending energy, one can minimize the energy density augmented by the constraint on the null Gaussian curvature through a Lagrange multiplier. 

The Euler-Lagrange equations associated with the bending energy in~\eqref{newenergy} and the natural forms derived in the Section \emph{Natural Forms of Shells} are
\[
\begin{aligned}
&2\Bigl[b_1^1 - \Bigl(\frac{1}{\Lambda_\textup{o}R}+\bar{\kappa}\Bigr)\Bigr] + 2\nu(b_2^2-\bar{\kappa}) - \lambda b_2^2=0\,,\\
&2(b_2^2-\bar{\kappa})+2\nu\Bigl[b_1^1-\Bigl(\frac{1}{\Lambda_\textup{o}R}+\bar{\kappa}\Bigr)\Bigr]- \lambda b_1^1= 0\,,
\end{aligned}
\] 
where~$\lambda$ is the Lagrange multiplier associated with the constraint on the Gaussian curvature. The equations admit the two solutions:
\begin{equation}
\begin{aligned}
b_1^1&=\frac{1}{\Lambda_\textup{o}R}+(1+\nu)\bar{\kappa}\,,\quad b_2^2=0\,,\quad\textup{with}\quad\overline{\mathcal{U}}_\textup{b}^1=(1-\nu^2)(1+ \bar{\kappa}\Lambda_\textup{o}R)^2\,,\\
b_1^1&=0\,,\quad b_2^2=\frac{\nu}{\Lambda_\textup{o}R}+(1+\nu)\bar{\kappa}\,,\quad\textup{with}\quad\overline{\mathcal{U}}_\textup{b}^2=(1-\nu^2)\bar{\kappa}^2\Lambda_\textup{o}^2R^2\,.\\
\end{aligned}
\end{equation}
The first solution corresponds to the case in which the cylinders bend by keeping their generatrices orientated in the same direction, while the second solution corresponds to a rotated shape. By equating the two energies of the two solutions we find a critical \emph{snapping} natural curvature that sets the threshold between the two
\begin{equation}
\bar{\kappa}_\textup{s}=-\frac{1}{2\Lambda_\textup{o}R}\,,
\end{equation} 
which is independent of the Poisson ratio. These results may be expressed concisely in terms of the (dimensionless) mean curvature as
\begin{equation}\label{34cyl}
Hh=\frac{hf(\bar{\kappa})}{2\Lambda_\textup{o}R}+\frac{1+\nu}{2}\bar{\kappa}h\,,\quad 
f(\bar{\kappa})=
\begin{cases}
1\,,\ \bar{\kappa}>-\frac{1}{2\Lambda_\textup{o}R}\,,\\
\nu\,,\ \bar{\kappa}<-\frac{1}{2\Lambda_\textup{o}R}\,.
\end{cases}
\end{equation}
Notice that when~$R\rightarrow\infty$ the cylindrical shell becomes a flat plate and equation~\eqref{34cyl} converges to that presented in~\cite{Pezzulla2016} except for a factor~$1/\Lambda_\textup{o}^2$, as given by equation~\eqref{34plate} (with~$\nu=1/2$). 

\begin{figure}[t]
\centering

\includegraphics[scale=1]{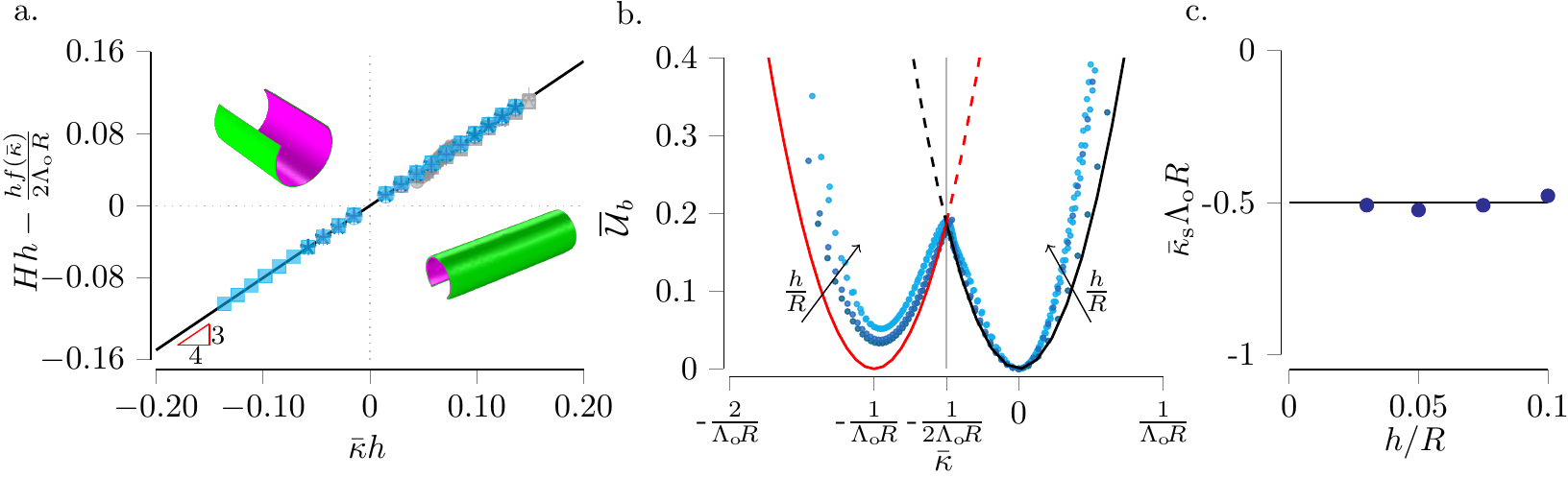}
	
\caption{Isometric bending of cylinders. (a) The theoretical prediction for the mean curvature (solid line) captures the numerical results for cylinders (blue symbols) as well as the numerical and experimental results (grey symbols) for the bending of bilayer plates from~\cite{Pezzulla2016}. (b) Energy profiles from theory (solid lines) and three-dimensional numerical simulations (symbols) for~$h/R=(0.02,0.04,0.08)$. (c) Dimensionless snapping natural curvature from theory (solid line) and three-dimensional numerical simulations (symbols).}
\label{fig:cylinder}
\end{figure}

Figure~\ref{fig:cylinder}~(a) shows the agreement of our theoretical prediction~\eqref{34cyl} with the numerical results for flat plates (from~\cite{Pezzulla2016}) and cylinders (blue symbols), and experimental results (grey symbols) from~\cite{Pezzulla2016} when~$\nu=1/2$. Notice how the \emph{snapped} shapes obtained for~$\bar{\kappa}<-1/(2\Lambda_\textup{o}R)$ are cylinders with generators rotated by~$90$ degrees with respect to the initial shape. Numerical simulations were performed to solve the geometrical problem within the context of finite (incompatible) tridimensional elasticity with large distortions using a neo-Hookean incompressible material model~\cite{Lucantonio2014} implemented in the commercial software COMSOL Multiphysics. The shells were made of two layers: one layer was subjected to a distortion field~$\vett{F}_\textup{o}=\lambda(\vett{e}_1\otimes\vett{e}_1+\vett{e}_2\otimes\vett{e}_2)+\vett{e}_3\otimes\vett{e}_3$, whereas the other one was subjected to~$\vett{F}_\textup{o}=\vett{I}$. Then, the geometry was reduced to that of the beam, and we measured the curvature of the beam for each~$\lambda$ to determine~$\bar{\kappa}$. Figure~\ref{fig:cylinder}~(b) shows the energy profiles~$\overline{\mathcal{U}}_\textup{b}^1$ (black line) and~$\overline{\mathcal{U}}_\textup{b}^2$ (red line) as functions of~$\bar{\kappa}$. Symbols represent energy profiles from three-dimensional numerical simulations that get closer to theoretical prediction for thinner shells. Notice also the emergence of a second minimum at zero energy, corresponding to a perfectly inverted and rotated cylinder, which becomes possible due to growth as the natural configuration continuously changes with~$\bar{\kappa}$. Finally, figure~\ref{fig:cylinder}~(c) shows the comparison between our theoretical prediction for the snapping natural curvature (solid line) and the numerical three-dimensional results. 

\subsection{Cones} Let us now consider a cone of length~$l$ with largest and smallest radius~$R_\textup{o}$ and~$R_\textup{l}$, respectively. In this case the metric is still flat but non homogeneous. In cylindrical coordinates~$(\eta^1,\eta^2)=(\theta,z)$, we have:
\begin{equation}
\bar{a}_{\alpha\beta}=
\Lambda_\textup{o}^2
\begin{pmatrix}
R(z)^2&0\\
0&1+c^2\\
\end{pmatrix}\,,\quad
\bar{b}_\alpha^{\eta}=
\begin{pmatrix}
\frac{1}{\Lambda_\textup{o}R(z)\sqrt{1+c^2}}+\bar{\kappa}&0\\
0&\bar{\kappa}\\
\end{pmatrix}\,,
\end{equation}
where~$R(z)=cz+R_\textup{o}$, with~$c=(R_\textup{l}-R_\textup{o})/l$. Again, we look for an isometric deformation of the cone by setting the stretching energy equal to zero. However, homogeneous second fundamental forms are not solutions of the Gauss-Codazzi-Mainardi equations:
\begin{equation}\label{GCMcones}
\begin{aligned}
b_{11},_z-b_{12},_\theta&=\frac{c}{R(z)}b_{11}+\frac{c}{1+c^2}R(z)b_{22}\,,\\
b_{12},_z-b_{22},_\theta&=-\frac{c}{R(z)}b_{12}\,,\\
b_{11}b_{22}-b_{12}^2&=0\,.
\end{aligned}
\end{equation} 
The bending energy has to be minimized under these differential and algebraic constraints. This can be performed in COMSOL Multiphysics, in which we minimize the bending energy augmented by the three constraints imposed through three Lagrange multipliers fields. Moreover, we can assume that symmetry is preserved during the deformation so that~$(),_\theta$ and~$b_{12}=b_{22}=0$. Consequently, two of the original Gauss-Codazzi-Mainardi equations are trivially satisfied, while the other one can be solved analytically as

\begin{equation}
b_{11},_z=\frac{c}{cz+R_\textup{o}}b_{11}\,\Rightarrow\, b_{11}(z)=A(cz+R_\textup{o})\,,\quad\! \! A\in\mathbb{R}\,.
\end{equation}
Substituting this solution in the bending energy density, this can be integrated to provide the bending energy up to the scalar parameter~$A$. Therefore, the energy is now just a function that can be minimized with respect to~$A$ and the results may be expressed in terms of the dimensionless mean curvature as

\begin{equation}\label{Hcones}
H(z)h=\frac{h}{2(cz+R_\textup{o})}\biggl[\frac{1}{\Lambda_\textup{o}\sqrt{1+c^2}}+\frac{(1+\nu)cl\bar{\kappa}}{\log(cl/R_\textup{o}+1)}\biggr]\,.
\end{equation}
Note that for~$c\rightarrow0$, the cone approaches a cylinder of radius~$R_\textup{o}$, and, as expected, equation~\eqref{Hcones} approaches~\eqref{34cyl} for cylinders with~$\bar{\kappa}>-1/(2R_\textup{o})$. Moreover, letting in addition~$R\rightarrow\infty$, we recover the result for flat plates. The result for cones, equation~\eqref{Hcones}, thus constitutes a generic result that holds for cones, cylinders, and plates. To make this more evident, we define

\begin{equation}
H_{\bar{\kappa}}(cl/R_\textup{o}):=\partial_{\bar{\kappa}}H(z)\frac{cz+R_\textup{o}}{R_\textup{o}}=\frac{(1+\nu)cl/R_\textup{o}}{2\log{(1+cl/R_\textup{o})}}\,,  
\end{equation}
which will prove to be useful to describe the influence of the slope of the cone on the deformed mean curvature. Notice indeed that for~$c=0$, $H_{\bar{\kappa}}=(1+\nu)/2$ that is exactly the first derivative of the mean curvature with respect to the natural curvature for plates and cylinders.

\begin{figure}[t]
\centering
\vspace{-0.5cm}
\includegraphics[scale=1]{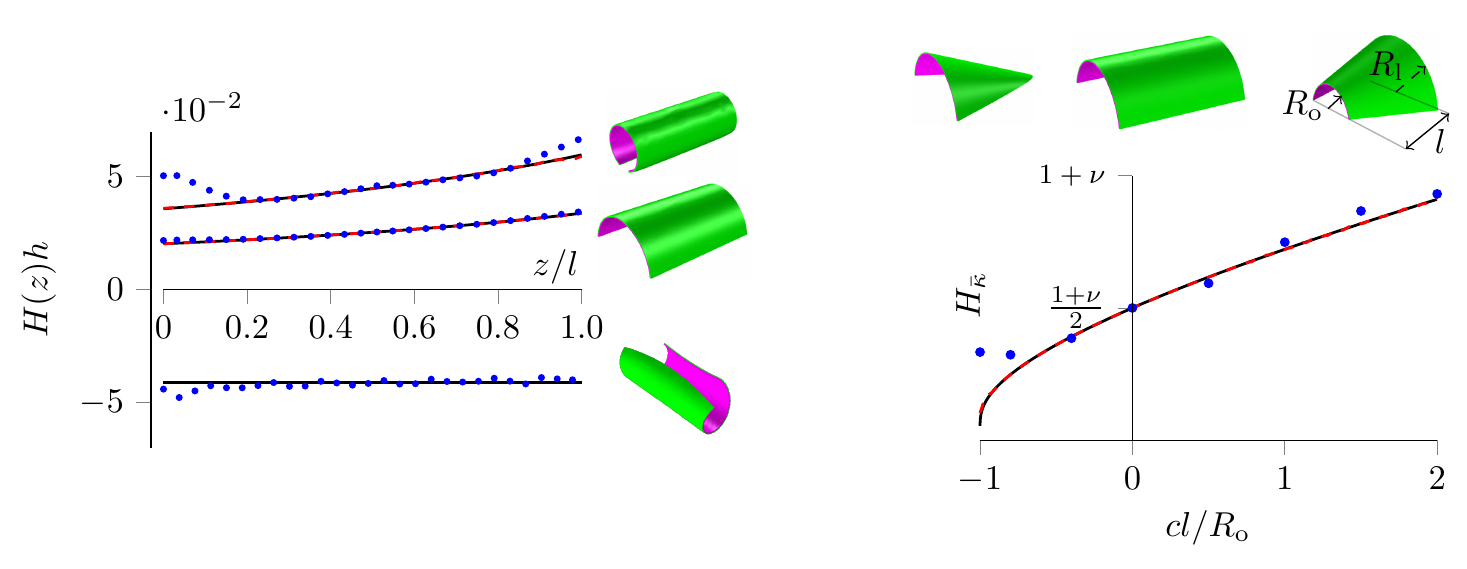}
\caption{Isometric bending of cones. Dimensionless mean curvature as a function of~$z/l$ for three different values of natural curvature~$\bar{\kappa}R_\textup{o}=(0.07,0.73,-1.43)$ and~$cl/R_\textup{o}=-0.4$ from the analytical model (solid line), numerical constrained minimization (dashed line) and 3D numerical results (symbols), and corresponding deformed shapes (left). Plot of~$H_{\bar{\kappa}}$ versus~$cl/R_\textup{o}$ from the analytical model (solid line), numerical constrained minimization (dashed line) and 3D numerical results (symbols). Three different cone geometries are shown for~$cl/R_\textup{o}=-1,0,1$, from left to right (right).}
\label{fig:cone}
\end{figure}
Figure~\ref{fig:cone}~(left) shows the comparison among 3D numerical results (symbols), numerical constrained minimization (dashed red line) and analytical model (solid lines) in terms of the dimensionless mean curvature versus the longitudinal coordinate~$z/l$ for cones. As the natural curvature~$\bar{\kappa}$ increases, the cone bends and both its two largest and smallest circles of radii~$R_\textup{o}$ and~$R_\textup{l}$ shrink, and the normalized mean curvature increases. While the analytical and constrained minimization models agree very well with each other, the 3D model shows the emergence of boundary layers that is a finite thickness effect. Figure~\ref{fig:cone} (right) shows how~$H_{\bar{\kappa}}$ varies with~$cl/R_\textup{o}$ according to the analytical prediction (solid line), numerical constrained minimization (dashed line) and 3D finite element model (symbols). The three cones above the plot correspond to~$cl/R_\textup{o}=-1,0,1$, from left to right: the cone corresponding to~$cl/R_\textup{o}=-1$ has a vertex at one end that is a pure tridimensional region of the body, completely ignored by our models. Indeed, while our analytical model predicts~$H_{\bar{\kappa}}=0$ (inconsistent with the geometry of the cone), the three-dimensional finite element results show a value greater than zero. 

As investigated for cylindrical shells, we studied the snapping of conical shells triggered by a decreasing natural curvature and noticed a snapping transition where the orientation of the cone rotates by~$90$ degrees. Surprisingly, the resulting shape after snapping is very close to a cylinder for~$|cl/R_\textup{o}|<1$, characterized by a roughly constant mean curvature along~$z$, as indicated by 3D numerical results (symbols). Furthermore, we applied the formula for the morphing of cylinders having a radius equal to~$R_\textup{o}$ (solid line), and found excellent agreement with numerics when~$|cR_\textup{o}/l|\ll1$ (see {\it Appendix}).  

\section{Doubly curved shells}
\label{DouShell}
Finally, we investigate the growth of a spherical shell, which is both intrinsically and extrinsically curved. If we denote by~$R$ the radius of the shell, we have in spherical coordinates~($u$,$v$)

\begin{equation}\label{natuformspollen}
\bar{a}_{\alpha\beta}=
\Lambda_\textup{o}^2
\begin{pmatrix}
R^2&0\\
0&R^2\sin^2u\\
\end{pmatrix}\,,\quad
\bar{b}_\alpha^{\eta}=
\begin{pmatrix}
\frac{1}{\Lambda_\textup{o}R}+\bar{\kappa}&0\\
0&\frac{1}{\Lambda_\textup{o}R}+\bar{\kappa}\\
\end{pmatrix}\,,
\end{equation}
where~$u\in[0,\theta]$ represents the colatitude of the shell ($\theta$ is called half-angle), while~$v\in[0,2\pi)$ is the azimuthal angle.
\begin{figure}[t]
\centering

\includegraphics[scale=1]{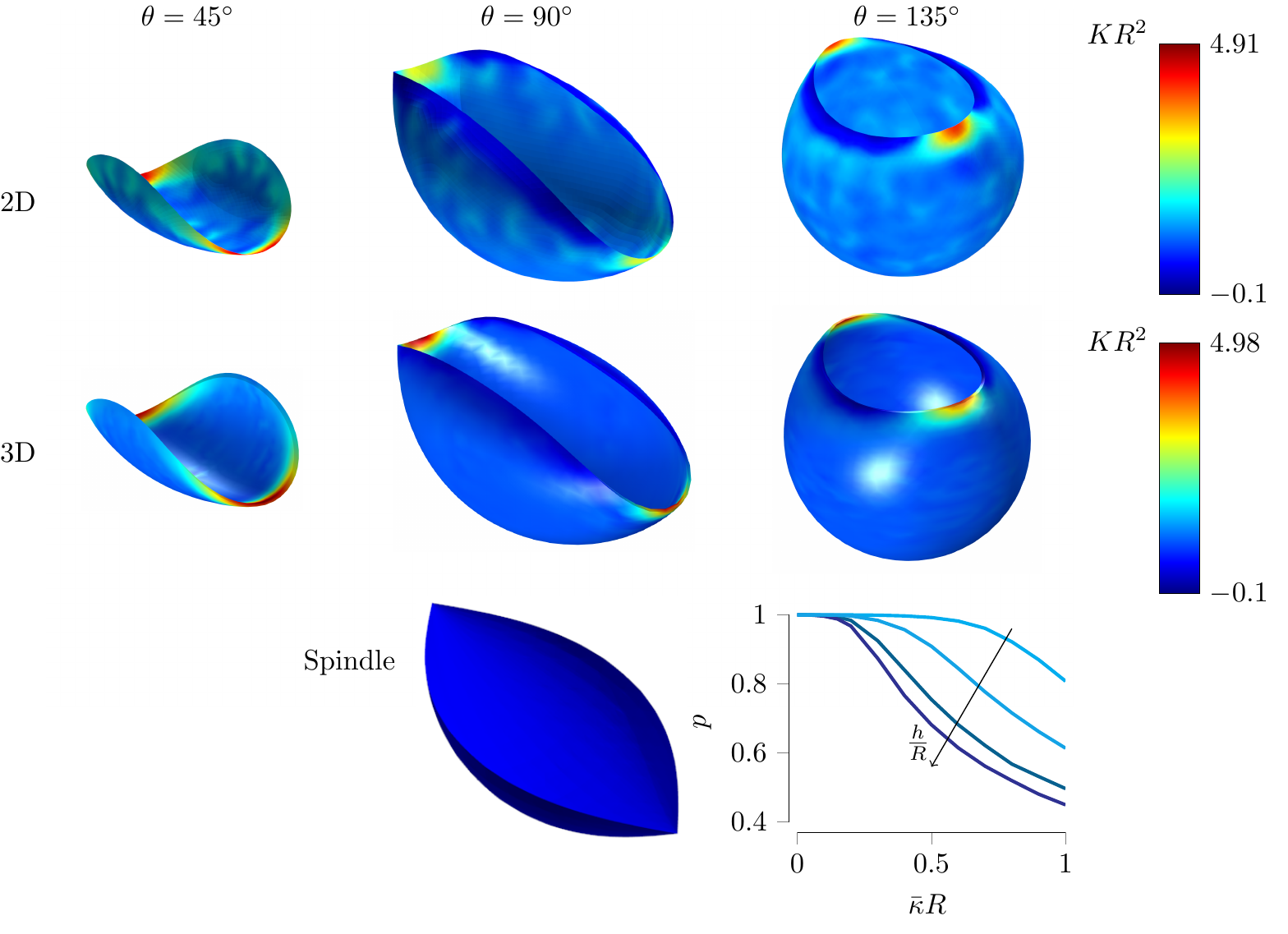}
	
\caption{Nearly-isometric deformations of a spherical shell under increasing natural curvature. Comparison of the two-dimensional model (first row) and the three-dimensional model (second row) for three different half-angles, at~$\bar{\kappa}R=0.99$. In the third row, the spindle corresponding to the minimization of the bending energy when neglecting boundary layers of width~$\sqrt{h/R}$ is shown. The plot shows the influence of the width of the boundary layer on the shape via the parameter~$p$.}
\label{fig:pollens}
\end{figure}
This configuration is analogous to the one that triggers harmomegathy in pollen grains~\cite{Katifori2010,Couturier2013}. To test the model presented in Section~\emph{Natural Forms of Shells}, we start by a comparison between three-dimensional numerical results based on the model presented in~\cite{Lucantonio2014}, and a two-dimensional model based based on $C^1$-continuous subdivision finite elements (SDFEs)~\cite{Cirak2001}. In the former, the outer layer of the shell swells while the inner layer shrinks, and we measure the corresponding natural curvature by running simulations of a beam; in the latter, equations~\eqref{abar} and~\eqref{bbar} are directly implemented. As a consequence, if the two models show two shapes with equal deformation (and curvatures) for the same amount of natural curvature, the model presented in Section \emph{Natural Forms of Shells} is verified. In figure~\ref{fig:pollens}, we present the deformed shapes of spherical shells as obtained from the two-dimensional model (first row) and the three-dimensional model (second row) for different values of the half-angle~$\theta=45^\circ, 90^\circ, 135^\circ$; color code represents the dimensionless Gaussian curvature~$KR^2$. The models provide shapes that are in excellent agreement with each other, being the differences in the Gaussian curvature smaller than~$1.4\%$. The boundary layer that develops along the edge clearly shows two peaks of the Gaussian curvature at two opposite locations. 

Contrary to the previous cases of cylinders and cones, there are no smooth surfaces that are isometric to a sphere (except any open subset of the sphere, which are of no interest for the problem at hand). A non smooth surface with homogenous positive Gaussian curvature is the spindle, and contains two vertices (singularities) at two opposite locations, exactly as the two peaks in the Gaussian curvature shown in the numerical results. A spindle can be parametrized (up to a scale factor) as a surface of revolution

\begin{equation}\label{spindle}
\vett{r}(\eta^1,\eta^2)=(\Phi(\eta^2)\cos(\eta^1),\Phi(\eta^2)\sin(\eta^1),\Psi(\eta^2))\,,
\end{equation}
where~$\Phi(\eta^2)=p\cos(\eta^2)$ and~$\Psi(\eta^2)=E(\eta^2\ \lvert\ p^2)$, where~$E$ denotes the incomplete elliptic integral of the second kind, and~$p$ is a parameter equals to~$1$ for a sphere and less than~$1$ for spindles: the smaller this parameter, the more elongated the shape of the spindle is. The Gaussian curvature is homogenous but singular at~$\eta^2=\pm\pi/2$ whereas the mean curvature is equal to
\begin{equation}
H(\eta^2)=\frac{(1+p^2\cos(2\eta^2))\sec(\eta^2)}{2p\sqrt{1-p^2\sin^2(\eta^2)}}\,,
\end{equation}
as well singular at~$\eta^2=\pm\pi/2$. The bending energy related to this configuration with respect to the natural forms~\eqref{natuformspollen} is then singular and diverges as

\begin{equation}
\overline{\mathcal{U}}_\textup{b}\sim\frac{p^2-1}{2p}\log{\Bigl(\frac{\pi}{2}-\eta^2\Bigr)}\,, \quad\text{as}\quad\eta^2\rightarrow\pi/2^-\,,
\end{equation}
analogously to the divergence of the bending energy of $e$--cones~\cite{Muller2008}. For~$p=1$, the spindle becomes a sphere and the singularities disappear. Therefore, the mapping~\eqref{spindle} can be interpreted as a family of deformations from a hemisphere to spindles. By neglecting an area around the two singularities with a magnitude comparable to the one of the boundary layer~$\sqrt{h/R}$, the bending energy becomes finite and can be minimized with respect to~$p$ to give the shape in the last row of figure~\ref{fig:pollens}. Contrary to what shown for $e$~cones in~\cite{Muller2008}, the spindle that minimizes the bending energy does depend on the width of boundary layer neglected in the integration of the energy, as shown in the plot of figure~\ref{fig:pollens}. We note that there is no singularity apparent in the realized shape of the shell, as opposed to the presence of the vertex in an $e$--cone. The reasons for this difference could be rooted in the geometry of surfaces with singularities, as the spindle is an elliptic surface while the cone is parabolic, however, this question is beyond the scope of this work.

Finally, as for cylinders and cones, a negative natural curvature can trigger snap-through instabilities in spherical shells. A shell with an increasing natural curvature will then buckle in spindle-like shapes, whereas shells with decreasing natural curvature will snap. The competition between snapping and buckling of spherical shells will be addressed in a future work.

\section{Conclusion} 

In this paper, we have studied the growth of shells induced by variations in natural curvature starting from the theory of non-Euclidean shells. A three-dimensional, non--mechanical stimulus is reduced to the natural first and second fundamental forms of the mid-surface of the structure, via a geometric criterion based on the linear projection of the reference metric. The effect of the stimulus results to be additive to the initial geometric structure except for a spherical stretching factor that accounts for homotheties.

We have applied the model to different thin structures, from flat plates to spherical shells, finding excellent agreement between theory and numerics. As the results are derived from a purely geometrical model, they are applicable to a large variety of stimuli and, as such, are general and scalable.

\appendix*

\section{Snapping and rotation for cones}

Similarly to cylinders, cones can snap and rotate under a negative (decreasing) natural curvature. However, differently from cylinders, an isometric snapping and rotation is forbidden by geometrical compatibility conditions. Indeed, the Gauss-Codazzi-Mainardi equations~\eqref{GCMcones} do not admit the snapped and rotated solution~$b_{11}=b_{12}=0$ and~$b_{22}\ne0$ since this implies
\begin{equation}\label{GCMc}
\frac{c}{1+c^2}R(z)b_{22}=0\,.
\end{equation}
This means that when the cone snaps and rotates as observed in the numerical simulations, the mid-surface stretches. However, when~$cl/R_\textup{o}\ll1$, the amount of stretch is small since equation~\eqref{GCMc} is satisfied at leading order. This explains why cones with small slopes snap and rotate very similarly to cylinders. 
\begin{figure}[h]

\centering

\includegraphics[scale=1]{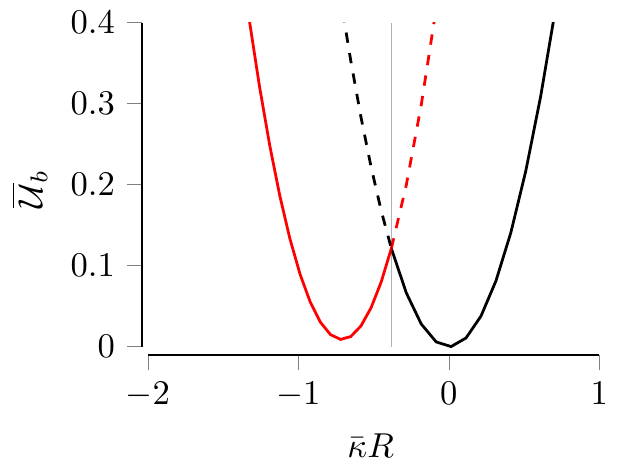}

\caption{Energy profiles for cone in the bending case (black line) and snapping/rotation case (red line).}
\label{conesappendix}
\end{figure}

To show why a rotated configuration is preferred for negative natural curvature also for conical geometries, we write the bending energy of cones:
\begin{equation}
\overline{\mathcal{U}}_\textup{b}=\int_\omega (1-\nu)\Bigl[\Bigl(\frac{A}{R(z)}-\Bigl(\frac{1}{\Lambda_\textup{o}R(z)\sqrt{1+c^2}}+\bar{\kappa}\Bigr)\Bigr)^2+(b_2^2-\bar{\kappa})^2\Bigr]+\nu\Bigl(\frac{A}{R(z)}-\frac{1}{\Lambda_\textup{o}R(z)\sqrt{1+c^2}}+b_2^2-2\bar{\kappa}\Bigr)^2d\omega\,,
\end{equation}
where we already used compatibility. This energy can be minimized with respect to~$A$ assuming~$b_2^2=0$ to get equation~\eqref{Hcones}. Similarly, when~$cl/R_\textup{o}\ll1$, it can also be minimized with respect to~$b_2^2$ assuming~$A=0$ to get
\begin{equation}\label{conerotated}
Hh=\frac{h}{2}\Bigl[(1+\nu)\bar{\kappa}+\frac{\nu}{\Lambda_\textup{o}cl\sqrt{1+c^2}}\log\Bigl(1+\frac{cl}{R_\textup{o}}\Bigr)\Bigr]=\frac{\nu h}{2\Lambda_\textup{o}R_\textup{o}}+\frac{1+\nu}{2}\bar{\kappa}h+O(c)\,,
\end{equation}
which is equal to the formula for the mean curvature of cylinders, at leading order.
The energies corresponding to the two different solutions~$(b_1^1\ne0,\ b_2^2=0)$, equation~\eqref{Hcones}, and~$(b_1^1=0,\ b_2^2\ne0)$, equation~\eqref{conerotated}, can be plotted as functions of~$\bar{\kappa}$ in figure~\ref{conesappendix}.
The black solid line corresponds to~$(b_1^1\ne0,\ b_2^2=0)$ while the red line corresponds to~$(b_1^1=0,\ b_2^2\ne0)$. The behavior is similar to that encountered for cylinders although is more complicated, and approximately valid only for~$cl/R_\textup{o}\ll1$. The critical value of the natural curvature is a function of the Poisson ratio in this case, and the geometry of the cone, while the second minimum of the bending energy is not a zero energy state.

\section*{Acknowledgments}
D.P.H. is grateful for financial support from the NSF CAREER CMMI--1454153.

\end{document}